\documentclass[12pt]{article}
  \usepackage{amsfonts,amssymb}
  \usepackage{tikz}
    \usetikzlibrary{arrows}
  \newtheorem{theorem}{Theorem}[section]
  \newtheorem{lemma}{Lemma}[section]
  \newtheorem{con}{Consequence}[section]
  \def\mflog{\mathop{\mathrm{log}}\nolimits}
  \def\mfln{\mathop{\mathrm{ln}}\nolimits}
  \def\mfdim{\mathop{\mathrm{dim}}\nolimits}
  \def\mfN{\mathop{\hbox{}\mathrm{N}}\nolimits}
  \def\mfd{\mathrm{d}}
\begin{document}
  \relax
    \title{Asymptotic estimate for the number of Gaussian packets on three
      decorated graphs}
    \author{%
      V.\,L. Chernyshev\footnote
        {National Research University Higher School of Economics (HSE),
          Myasnitskaya Street, 20, Moscow, 101978, Russia;
        vchern@mech.math.msu.su},
      A.\,A. Tolchennikov\footnote
        {A. Ishlinsky Institute for Problems in Mechanics, Russian Academy of Sciences,
          Pr. Vernadskogo, 101-1, Moscow, 119526.}}
    \date{02.03.2014}
    \maketitle
    \begin{abstract}
      We study a topological space obtained from a graph via replacing vertices with smooth
      Riemannian manifolds, i.e. a decorated graph. We construct a semiclassical
      asymptotics of the solutions of Cauchy problem for a time-dependent
      Schr\"{o}dinger equation on a decorated graph with a localized initial
      function. The main term of our asymptotic solution at an arbitrary
      finite time is the sum of Gaussian packets and generalized Gaussian
      packets. We study the number of such packets as time goes to infinity.
      We prove asymptotic estimations for this number for the following decorated
      graphs: cylinder with a segment, two dimensional torus with a segment, three dimensional torus with a segment.
      Also we prove general theorem about a manifold with a segment and apply it to the case of a uniformly secure manifold.
    \end{abstract}
  \relax
  \section{Introduction}
\noindent
  Differential equations and differential operators on decorated graphs
  have been intensively studied over the past thirty years (see, for
  example, \cite{Kuch_Berk_book}, \cite{BG}, \cite{FP}, \cite{Pav}  and
  references therein).
\par
\indent
  The study of the motion of Gaussian packets arises when considering a
  Cauchy problem for a time-dependent Schr\"{o}dinger equation. Let us
  consider the Cauchy problem for a decorated graph, which is a singular space obtained by gluing the
  ends of segments to the surfaces of dimension two or three. The initial conditions are a
  Gaussian packet with support on one of the edges. We look for a
  semiclassical solution (see, for example, \cite{cwkb}, \cite{mf} and
  references therein). Upon reaching the end of the segment, the packet
  forms an expanding wavefront on the surface. If the front reaches
  another point of gluing then a new Gaussian packet starts to move along the
  corresponding edge and so on. We are interested in finding the
  asymptotic behavior of the number of supports of the Gaussian packets on the
  edges of a decorated graph at time $T$. A detailed description of the
  formulation of this problem is given in \cite{trsa}.
\par
\indent
  Wavefront propagation on surfaces is associated with the properties of
  geodesics on the surface. There are two variations. In the first the
  number of geodesics connecting two points on the surface is finite. As
  an example we can take a standard sphere with an edge. In this situation
  we can construct from one-dimensional edges and geodesics on the
  surfaces an equivalent metric graph and describe the statistics of the
  distribution of Gaussian packets using our results obtained before (see,
  e.g., \cite{mian}, \cite{RJoMP}). Theorems on the asymptotic behavior of the number of
  packets $\mfN(T)$ and the uniformity of their distribution for this case
  were proven in \cite{trsa}.
\par
\indent
  The second variation is more general, where the equivalent
  geometric graph is infinite. The main question here is how the number of
  geodesics joining two given points increases as time goes to infinity. The time and the maximum length are synonymous in this context. Some results in this area
  can be found in \cite{Berger}.
  Let $h$ be a topological entropy
  for a compact Riemannian manifold $M$. R. Ma\~{n}\'{e} has shown in
  \cite{Mane} (see references therein) that for quite general situations
  $h=\lim\limits_{T\to\infty}\frac{1}{T}\int_{M{\times}M}\mflog CF_T(x,y)\mfd\mu(x)\mfd\mu(y)$,
  where $CF_T(x, y)$ is the number of geodesics joining
  $x$ to $y$ of maximum length $T$. But this equation may fail if $M$ has
  conjugate points. There are examples where the growth of $CF_T (x, x)$
  is arbitrarily large for some exceptional points $x \in M$ (see
  \cite{BuGu}). There are also examples in which the limit in the equation
  is smaller than the topological entropy for an open set of
  configurations (see \cite{BuPa2}).
\par
\indent
  A Riemannian manifold is said to be uniformly secure (see \cite{BuGu})
  if there is a finite number $k$ such that all geodesics connecting an
  arbitrary pair of points in the manifold can be blocked by $k$ point
  obstacles. It is proven in \cite{BuGu} that the number of geodesics with
  length $\leq T$ between every pair of points in a uniformly secure
  manifold grows polynomially as $T \to \infty$. According to the results of
  M. Gromov and R. Ma\~{n}\'{e} (see \cite{Mane}), the fundamental group of such a manifold
  is virtually nilpotent, and the topological entropy of its geodesic flow
  is zero. Furthermore, if a uniformly secure manifold has no conjugate
  points, then it is flat. This follows from the virtual nilpotency of its
  fundamental group either via the theorems of Croke-Schroeder and
  Burago-Ivanov, or by a more recent work of Lebedeva \cite{Leb}.
\par
\indent
  In the present article we consider the case where the number of
  geodesics grows polynomially, for example a compact
  Riemannian manifold that is uniformly secure. We study in detail two
  examples: a standard cylinder and a flat torus with one edge. In the
  situation where the equivalent lengths of the edges of the graph are
  linearly independent over the field of rational numbers, there is a
  one-to-one correspondence between the times in which the birth of new
  Gaussian packets occurs and nonnegative solutions of infinite linear
  inequality where the right hand side equals $T$. Such inequalities arise
  in number theory, namely in the analysis of the asymptotic behavior
  of the number of partitions of natural numbers. We can recall the
  results of G.\,H. Hardy and S. Ramanujan, J.\,V. Uspensky, G.
  Rademacher, P. Erd\"{o}s and others. In our case, the length of the
  geodesics will not be a sequence of natural numbers. Therefore, at the first
  stage of our research we obtain only an upper bound from the classical
  results. On the other hand, a lower bound was proven in \cite{trsa},
  namely for infinite number of edges $\mfN(T)$ grows faster than any polynomial. Later
  we found a connection between the problem we had studied and questions
  arising in the study of Bose-Maslov gas entropy, which in recent years
  was carried out by V.\,P. Maslov and V.\,E. Nazaikinskii. We used a 2013
  result of V.\,E. Nazaikinskii (see \cite{ve_n}) to
  obtain asymptotic formulas for the logarithm of the number of Gaussian
  packets on the edge. If linear independence over $\mathbb{Q}$ does
  not hold (such situation is certainly possible because we can even have
  many geodesics with the same length, see e.g. \cite{Marklof}), the
  resulting asymptotic formula becomes an upper bound for $\mfN(T)$.
  It should be noted that the same results can be obtained using Additive Abstract Prime Number Theorem from \cite{Knopfmacher}.
   A computer experiment conducted jointly with O.\,V. Sobolev has confirmed the
  correctness of the estimates. The simulation results for a decorated graph
  obtained by attaching a length of the cylinder are also given in \cite{trsa}.
\par

    \subsection{Preliminary remarks and definitions}
\noindent
  A {\itshape decorated graph} is a topological space, obtained from a metric
  graph by replacing vertices by smooth manifolds of dimensions two or
  three. Consider a finite number of smooth complete Riemannian manifolds
  $M_j$, and a number of segments $\gamma_i$, endowed with regular
  parametrization. For each endpoint $y$ of an arbitrary segment
  $\gamma_j$ fix a point $\tilde y$ on one of the manifolds $M_k$; we
  assume all points $\tilde y$ to be distinct. A decorated graph
  $\Gamma_d$ is a quotient space of the disjoint sum $\bigsqcup_k
  M_k\bigsqcup_j\gamma_j$ by the equivalence $y \sim \tilde y$.
\par
\indent
  The Schr\"{o}dinger equation on a decorated graph is defined as follows
  (see \cite{BG} and \cite{trsa} for detailed explanation; the original
  ideas were presented in \cite{FP},\cite{Pav}).
\par
\indent
  Let $V$ be a real valued continuous function on $ \Gamma_d$, smooth on
  the edges. Let $V_j$ and $V_k$ be restrictions of $V$ to $\gamma_j$ and
  to $M_k$ respectively. Consider a direct sum $
  \widehat{H}_0=\bigoplus\limits_{j=1}^{E}\left(-\frac{h^2}{2}\frac{\mfd^2}{\mfd
  z_j^2}+V_j\right)\bigoplus\limits_{k=1}^{V}\left(-\frac{h^2}{2}\Delta_k+
  V_k\right) $ with the domain
  $H^2(\Gamma)=\bigoplus\limits_{j=1}^{E}H^2(\gamma_j)\bigoplus\limits_{k=
  1}^{V} H^2(M_k).$ Here $\frac{\mfd^2}{\mfd z_j^2}$ is an operator of the second
  derivative on $\gamma_j$ with respect to a fixed parametrization with
  Neumann boundary conditions, $\Delta_k$ is the Laplace--Beltrami
  operator on $M_k$.
\par
\indent
  {\bfseries Definition.} The {\itshape Schr\"{o}dinger operator} $\widehat{H}$ is a
  self-adjoint extension of the restriction $\widehat{H_0}|_{L}$, where $
  L=\{\psi\in H^2(\Gamma),\quad \psi(y_s)=0\}.$
\par
\indent
  Domain of the operator $\widehat{H}$ contains functions with
  singularities in the points $y_j$. Namely, let $G(x,y,\lambda)$ be the
  Green function on $M_k$ (integral kernel of the resolvent) of $\Delta$,
  corresponding to the spectral parameter $\lambda$. This function has the
  following asymptotics as $x\to y$: $G(x,y,\lambda)=F_0(x,y)+F_1$, where
  $F_1$ is a continuous function and $F_0$ is independent of $\lambda$ and
  has the form $F_0=-\frac{c_2}{2\pi}\mfln\rho$ if $\mfdim M=2$
  and $F_0=\frac{c_3}{4\pi\rho}$ if $\mfdim M=3$. Here $c_j(x,y)$ is
  continuous, $c_j(y,y)=1$, $\rho$ is the distance between $x$ and $y$.
  The function $\psi$ from the domain of the operator $\widehat{H}$ has
  the following asymptotics as $x\to y_j$: $ \psi=\alpha_jF_0(x)+b_j+o(1)$.
\par
\indent
  Now for each endpoint of the segment consider a pair $\psi(y)$,
  $h\psi^\prime(y)$ and a vector $\xi=(u,v)$,\hfil\penalty-10000
  \hbox{}\kern4em
    $u=(h\psi^\prime(y_1),\dots,h\psi^\prime(y_{2E}),\alpha_1,\dots,\alpha_{2E})$,\hfil\penalty-10000
  \hbox{}\kern4em
    $v=(\psi(y_1),\dots,\psi(y_{2E}),hb_1,\dots,hb_{2E})$.\hfil\penalty-10000
  Consider a standard skew-Hermitian form
  $[\xi^1,\xi^2]=\sum_{j=1}^{4E}(u^1_j\bar v^2_j-v_j^1\bar u_j^2)$ in
  $\mathbb{C}^{4E}\oplus\mathbb{C}^{4E}$. Let us fix the Lagrangian plane
  $\Lambda\subset \mathbb{C}^{4E}\oplus\mathbb{C}^{4E}$. A self-adjoint
  extension $\widehat{H}$ is defined by the coupling conditions $\xi\in
  \Lambda$ or equivalently $\quad -i(I+U)u+(I-U)v=0,$ where $U$ is a
  unitary matrix defining $\Lambda$ and $I$ is an identity matrix. We will
  consider only local coupling conditions, i,e.
  $\Lambda=\bigoplus\limits_y{\Lambda}_y$, where ${\Lambda}_y\subset
  \mathbb C^4$ is defined for each point $y$ separately.
\par
\indent
  In Theorem 3.1 of the article \cite{trsa} it is shown how scattering of
  a Gaussian packet at the point of gluing of the edge to the surface
  takes place.
\par
\indent
  Further analysis of the number of packets on a decorated graph is as
  follows. Since wavefront propagation occurs along the geodesic, we can
  construct a matching metric graph for a given decorated graph. Vertices
  for this new graph are the points of gluing and edges are all geodesics
  on the surfaces connecting points of gluing and the old edges. We are
  interested in the number of old Gaussian packets, i.e. those whose
  support will lie on the edges of the original graph.
\par
\indent
  The number of packets could change only in those moments of time that
  have the form of linear combinations of edge travel time. So the number of
  packets is equal to the number of sets $\{n_j\}$ satisfying some
  inequations of this kind:
  $$
    n_1t_{l_1}+\ldots+n_mt_{l_m}{\leq}T,
    \label{ineq}
  $$
  where $t_j$ is a travel time of the $j$-th edge of
  equivalent graph and all $t_j$ are linearly independent over
  $\mathbb{Q}$. In the case where an equivalent graph contains an infinite
  number of edges we obtain an infinite inequality.
\par
\indent
  In the work \cite{ve_n} of V.\,E.Nazaikinskii the formula for $\mfln \mfN(E)$
  i.e. entropy of a gas with full energy less than $E$ was obtained. $\mfN(E)$
  is defined via the number of solutions of an infinite linear inequality,
  where $E$ is on the right hand side of the inequality.
\par
\indent
  Let us state this useful theorem.
\par
\begin{theorem}
  Let $\mfN(T)$ be the number of non-negative integer solutions of inequality
  $\sum_{i=1}^\infty \lambda_i N_i \le T$ and for sequence $\lambda_j$
  a counting function $\rho(\lambda) = \#\{j | \lambda_j \le \lambda\}$ has
  asymptotics
  $$
    \rho(\lambda) = c_0 \lambda^{1+\gamma} (1 + O(\lambda^{-\varepsilon})), \varepsilon > 0.
  $$
  Then
  $$
    \mfln \mfN(T) =
    (\gamma + 2) \left( \frac{c_0 \Gamma (\gamma + 2) \zeta(\gamma +
    2)}{(\gamma + 1)^{\gamma + 1}} \right)^{\frac{1}{\gamma + 2}}
    T^{\frac{\gamma + 1}{\gamma + 2}} (1+o(1))
  $$
  as $T$ goes to infinity.
  \label{th1}
\end{theorem}
\indent
  In next two sections we will consider two examples which show how to
  construct the asymptotics for $\mfN(T)$ for a given decorated graph.
\par
    \relax
  \relax
  \section{Decorated graph, obtained by gluing a segment to a cylinder}
\noindent
  Let us consider a circular cylinder with a length of a circle equaling
  $b$. Points $A$ and $B$ lie on a ruling of the cylinder at the distance
  $a$ from each other. The wave front that begins to spread from the point
  $A$ will reach the point $A$ again in the time of the form $nb$. Point
  of the front which is the first to return to the point $A$ will pass the
  cycle $e'$ with the passage time $t'=b$. Similarly, we define a cycle
  $e''$ from $B$ to $B$ with the passage time $t''=b$.
\par
\indent
  Let us assume that wavefront propagates from the point $A$ reaches the
  point $B$ at time $\sqrt{(kb)^2+a^2} (k \ge 0)$. One of points on wave
  front that reaches $B$ will passed way $e_k$ during the time $t_k =
  \sqrt{(kb)^2+a^2}$. Also, let $e'''$ be a segment that is glued to the
  cylinder with the travel time $t'''$.
\par
\indent
We can compose an infinite graph from the edges obtained earlier (see Pic.\,1).\hfil\penalty-10000
\vbox{\fontsize{10pt}{10pt}\selectfont
  \hbox to\linewidth{\hss\begin{tikzpicture}
    \path[draw, color = red!75!black, line width = 1pt]
      (-70pt,0pt)
        .. controls (-30pt,25pt) and (-10pt,25pt) ..
      ( 0pt,25pt)
        .. controls (10pt,25pt) and (30pt,25pt) ..
      ( 70pt,0pt)
      (-70pt,0pt)
        --
      ( 70pt,0pt)
      (-70pt,0pt)
        .. controls (-30pt,-25pt) and (-10pt,-25pt) ..
      ( 0pt,-25pt)
        .. controls (10pt,-25pt) and (30pt,-25pt) ..
      ( 70pt,0pt)
      (-70pt,0pt)
        .. controls (-30pt,-50pt) and (-10pt,-50pt) ..
      ( 0pt,-50pt)
        .. controls (10pt,-50pt) and (30pt,-50pt) ..
      ( 70pt,0pt)
      {(0pt,-58pt) node [color = black] {$\vdots$}}
      (-70pt,0pt)
        .. controls (-30pt,-70pt) and (-10pt,-70pt) ..
      ( 0pt,-70pt)
        .. controls (10pt,-70pt) and (30pt,-70pt) ..
      ( 70pt,0pt)
      {(0pt,-76pt) node [color = black] {$\vdots$}}
      (-70pt,0pt)
        .. controls (-115pt,45pt) and (-115pt,-45pt) ..
      (-70pt,0pt)
      ( 70pt,0pt)
        .. controls ( 115pt,45pt) and ( 115pt,-45pt) ..
      ( 70pt,0pt)
    ;
    \path[draw, color = blue!75!black, line width = 1pt]
      (0pt,25pt) node [anchor = south] {$e'''$}
      (0pt,0pt) node [anchor = south] {$e_0$}
      (0pt,-25pt) node [anchor = south] {$e_1$}
      (0pt,-50pt) node [anchor = south] {$e_2$}
      (-70pt,5pt) node [anchor = south] {$A$}
      ( 70pt,5pt) node [anchor = south] {$B$}
      (-105pt,0pt) node [anchor = base east] {$e'$}
      ( 105pt,0pt) node [anchor = base west] {$e''$}
    ;
    \path[draw, fill = green!75!black, line width = 0pt]
      (-70pt,0pt) circle (2pt)
      ( 70pt,0pt) circle (2pt)
    ;
  \end{tikzpicture}\hss}
  \kern.5em
  \leftskip   = 0pt plus1fill
  \rightskip  = 0pt plus1fill
  \noindent Pic.\,1.  Equivalent graph.\par}
\par
\kern1em
\begin{theorem}
  For decorated graph obtained by attaching a segment to a flat cylinder, the
  following asymptotic estimate holds, as $T$ goes to infinity:
  $$
    \mfln \mfN(T) \le \sqrt \frac{2}{3b} \ \pi \ T^{\frac12} (1+o(1))
  $$
  If times $\{t'\} \cup \{t_i \}_{i=0}^\infty $ are linearly independent
  over $\mathbb{Q}$ then inequality turns into equality. For almost all real $a$ and $b$ times $\{t'\} \cup \{t_i \}_{i=0}^\infty $ are linearly independent
  over $\mathbb{Q}$.
  \label{th2}
\end{theorem}
\indent
  \textbf{Proof} We list all times at which $\mfN(T)$ grows by one. It happens when:
\par
\indent
  1) A packet arrives at point $A$. I.e. at time of a kind:
  $$
    T = t' n' + t'' n'' + t''' n''' + \sum_{i=0}^k t_i n_i
  $$
  for some $k$, and non-negative integer $n', n'', n_0, \ldots n_k$
  satisfy the following conditions:\hfil\penalty-10000
    \hbox{}\kern4em a) $\sum_{i=0}^k n_i + n'''$ is even\hfil\penalty-10000
    \hbox{}\kern4em b) if $\sum_{i=0}^k n_i + n''' = 0$, then $n'' = 0$,\hfil\penalty-10000
    \hbox{}\kern4em c) if $\sum_{i=0}^k n_i + n''' > 0$, then $n'' \ge 0$.
\par
\indent
  A set of paths $\gamma_T$ from $A$ to $A$ corresponds to every moment of
  time $T$.
\par
\indent
  Function $\mfN(T)$ does not increase at time $T$ if the following condition
  is fulfilled: there is a path in $\gamma_T$, by which we return to point
  $A$ and we arrive by $e'''$ at final moment of time. Obviously, there is
  no such path if and only if $n'''=0$.
\par
\indent
  Thus a new packet is born in point $A$ at times of a kind $t'
  n+\sum_{i=0}^k t_i n_i$, where $n_i\ge 0$, $n\ge 0$, $\sum_{i=0}^k n_i$
  is even.
\par
\indent
  2) A packet arrives at point $B$. Similarly, a new packet is born in
  point $B$ at times of a kind $t' n+\sum_{i=0}^k t_i n_i$, where $n_i\ge
  0$, $n\ge 0$, and $\sum_{i=0}^k n_i$ is odd.
\par
\indent
  Thus $\mfN(T)$ equals the number of times of a kind $t' n+\sum_{i=0}^k t_i
  n_i, n_i\ge 0, n\ge 0$, and times less than $T$ . If $\{t'\} \cup \{t_i
  \}_{i=0}^\infty $ are linearly independent over $\mathbb{Q}$, then there
  exists one-to-one correspondence between such times and sets
  $(n,n_0,n_1,...)$. Hence, $\mfN(T)$ is equal to the number of solutions of
  inequality $t' n+\sum_{i=0}^k t_i n_i \le T$. Let us use the theorem
  \ref{th1}: $\rho(\lambda) = [\frac{1}{b} \sqrt{\lambda^2 - a^2}]$, thus
  $c_0=\frac{1}{b}, \gamma=0$ and we obtain
  $$
    \mfln \mfN(T) = \sqrt \frac{2}{3b} \pi T^{\frac12} (1+o(1)).
  $$
  If $\{t'\} \cup \{t_i \}_{i=0}^\infty $ are linearly dependent over $\mathbb{Q}$, then
  different sets $(n,n_0,..)$ correspond to one moment of time and then
  $$
    \mfln \mfN(T) \le \sqrt \frac{2}{3b} \pi T^{\frac12} (1+o(1)).
  $$

It remains only to explain  because of why for almost all real $a$ and $b$ times $\{t'\} \cup \{t_k \}_{k=0}^\infty $ are linearly independent over $\mathbb{Q}$. Let us take $b=1$ without loss of generality. It is suffcient to prove that a set $TS=\cup_{k=0}^\infty\{t_k=\sqrt{k^2+a^2} \}$ is linearly independent over $\mathbb{Q}$. Suppose that this is not true. This means that there is a finite list of rational numbers $\alpha_j, j=1,\ldots,m$ such that finite linear combination of elements of $TS$ with coefficients $\alpha_j$ equals zero. Hence the set of $a$ values such that $TS$ is $\mathbb{Q}$-linearly dependent is smaller than the set of finite sequences of rational numbers. Since the latter set is countable, so is the set of such $a$. Hence the set $TS$ is linearly independent over $\mathbb{Q}$ for almost all real $a$.

{\bf Remark.} One can prove that times $t_k$ are $\mathbb{Q}$-linearly independent if $a$ is a transcendent number (for example one can take $a=\pi$) in the following manner.  Since $a$ is transcendent, hence $\mathbb{Q}(a)$ is isomorphic to the field $\mathbb{Q}(x)$ of rational functions over $\mathbb{Q}$. To show that the numbers $\sqrt{n^2+a^2}$ are $\mathbb{Q}$-linearly independent is therefore equivalent to the statement that the functions $\sqrt{n^2+x^2}$ are linearly independent over $\mathbb{Q}$. Suppose that there exists a nontrivial linear combination of such functions that equals zero: $\sum\limits_{j=1}^n\alpha_j\sqrt{j^2+x^2}=0$. We will prove that all $\alpha_j=0$ in the following way. We take one summand $\alpha_j\sqrt{j^2+x^2}$ and place it on the right hand side. In the neighborhood of the point $j$ on the complex plane the rest of the sum is holomorphic function, so $\alpha_j$ should be zero. We can repeat this procedure for all $j$.

  \section{Decorated graph obtained by gluing segment to flat torus}
\noindent
  Let us take a flat torus with fundamental cycles of lengthes $a$ and
  $b$. Let us consider a fundamental rectangle with sides $a, b$ and take
  points $A=(0,0)$, $B=(c,d)$ in it. We glue a segment $e'''$ with travel
  time $t'''$ to points $A, B$.\hfil\penalty-10000
\vbox{\fontsize{10pt}{10pt}\selectfont
  \hbox to\linewidth{\hss\begin{tikzpicture}
    \path[draw, color = red!75!black, line width = 1pt]
      (-80pt,0pt)
      ++(0pt,  0pt) -- +(160pt,0pt)
      ++(0pt,-25pt) -- +(160pt,0pt)
      ++(0pt,-25pt) -- +(160pt,0pt)
      ++(0pt,-25pt) -- +(160pt,0pt)
      (-75pt,5pt)
      ++(0pt, 0pt) -- +(0pt,-85pt)
      ++(50pt, 0pt) -- +(0pt,-85pt)
      ++(50pt, 0pt) -- +(0pt,-85pt)
      ++(50pt, 0pt) -- +(0pt,-85pt)
    ;
    \path[draw, color = blue!75!black, line width = 0pt, fill = green!75!black]
      (-25pt,0pt) circle (2pt) node [anchor = south east] {$A$}
      ++(35pt,-20pt) circle (2pt) node [anchor = base west] {$B$}
    ;
    \begin{scope}[color = black, line width = .5pt, arrows = latex'-latex']
      \path[draw]
        (-25pt,3pt) -- +(50pt,0pt)
      ;
      \path[draw]
        (-28pt,0pt) -- +(0pt,-25pt)
      ;
      \path[draw]
        (-25pt,-20pt) -- +(35pt,0pt)
      ;
      \path[draw]
        (10pt,0pt) -- +(0pt,-20pt)
      ;
    \end{scope}
    \path[draw, color = black]
      (-25pt,3pt) +(25pt,0pt) node [anchor = south] {$b$}
      (-28pt,0pt) +(0pt,-12.5pt) node [anchor = east] {$a$}
      (-13pt,-20pt) node [anchor = south] {$c$}
      (10pt,-10pt) node [anchor = east] {$d$}
    ;
  \end{tikzpicture}\hss}
  \kern.75em
  \leftskip   = 0pt plus1fill
  \rightskip  = 0pt plus1fill
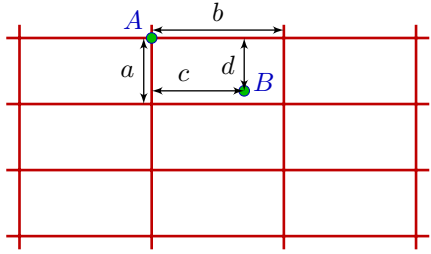
  \noindent Pic.\,2. Gluing of a segment to the flat torus.\par}
\par
\kern1em
\indent
  Wave packet begins its propagation from point $A$ and reaches point $A$
  again at times of the following kind $\{\sqrt{(na)^2+(mb)^2}| n\ge 0,
  m\ge 0, n^2+m^2\ne 0 \}$. Let us arrange them in ascending order: $t'_0,
  t'_1, \ldots$. One of the two front points reaches $A$ having pass the
  way $e'_k$ in time $t'_k$ (the other point passes this way in the other
  direction).
\par
\indent
  In a similar way, we consider front propagating from point $B$ and we
  obtain times $\{t''_i\}_{i=0}^\infty$ (where $t''_i=t'_i$) and paths
  $\{e''_i\}_{i=0}^\infty$.
\par
\indent
  Front propagating from point $A$ reaches point $B$ in times of the
  following kind $\{\sqrt{(c+na)^2+(d+mb)^2}| n,m\in \mathbb{Z} \}$. Let
  us arrange them in ascending order: $t_0, t_1, \ldots$. Let front point
  reach $B$ in time $t_k$ having pass way $e_k$.
\par
\indent
  We can construct an infinite graph from paths obtained before (see
  Pic.\,3).\hfil\penalty-10000
\vbox{\kern-2.5em\fontsize{10pt}{10pt}\selectfont
  \hbox to\linewidth{\hss\begin{tikzpicture}
    \path[draw, color = red!75!black, line width = 1pt]
      (-70pt,0pt)
        .. controls (-30pt,25pt) and (-10pt,25pt) ..
      ( 0pt,25pt)
        .. controls (10pt,25pt) and (30pt,25pt) ..
      ( 70pt,0pt)
      (-70pt,0pt)
        --
      ( 70pt,0pt)
      (-70pt,0pt)
        .. controls (-30pt,-25pt) and (-10pt,-25pt) ..
      ( 0pt,-25pt)
        .. controls (10pt,-25pt) and (30pt,-25pt) ..
      ( 70pt,0pt)
      (-70pt,0pt)
        .. controls (-30pt,-50pt) and (-10pt,-50pt) ..
      ( 0pt,-50pt)
        .. controls (10pt,-50pt) and (30pt,-50pt) ..
      ( 70pt,0pt)
      {(0pt,-58pt) node [color = black] {$\vdots$}}
      (-70pt,0pt)
        .. controls (-30pt,-70pt) and (-10pt,-70pt) ..
      ( 0pt,-70pt)
        .. controls (10pt,-70pt) and (30pt,-70pt) ..
      ( 70pt,0pt)
      {(0pt,-76pt) node [color = black] {$\vdots$}}
      (-70pt,0pt)
        .. controls (-100pt,25pt) and (-100pt,-25pt) ..
      (-70pt,0pt)
      (-70pt,0pt)
        .. controls (-120pt,45pt) and (-120pt,-45pt) ..
      (-70pt,0pt)
      (-70pt,0pt)
        .. controls (-150pt,75pt) and (-150pt,-75pt) ..
      (-70pt,0pt)
      ( 70pt,0pt)
        .. controls ( 100pt,25pt) and ( 100pt,-25pt) ..
      ( 70pt,0pt)
      ( 70pt,0pt)
        .. controls ( 120pt,45pt) and ( 120pt,-45pt) ..
      ( 70pt,0pt)
      ( 70pt,0pt)
        .. controls ( 150pt,75pt) and ( 150pt,-75pt) ..
      ( 70pt,0pt)
    ;
    \path[draw, color = blue!75!black, line width = 1pt]
      (0pt,25pt) node [anchor = south] {$e'''$}
      (0pt,0pt) node [anchor = south] {$e_0$}
      (0pt,-25pt) node [anchor = south] {$e_1$}
      (0pt,-50pt) node [anchor = south] {$e_2$}
      (-70pt,5pt) node [anchor = south] {$A$}
      ( 70pt,5pt) node [anchor = south] {$B$}
      (-90pt,0pt) node [anchor = base east] {$e_0'$}
      (-105pt,0pt) node [anchor = base east] {$e_1'$}
      {(-117pt,0pt) node [anchor = base east, color = black] {...}}
      ( 90pt,0pt) node [anchor = base west] {$e_0''$}
      ( 105pt,0pt) node [anchor = base west] {$e_1''$}
      {( 117pt,0pt) node [anchor = base west, color = black] {...}}
    ;
    \path[draw, fill = green!75!black, line width = 0pt]
      (-70pt,0pt) circle (2pt)
      ( 70pt,0pt) circle (2pt)
    ;
  \end{tikzpicture}\hss}
  \kern.5em
  \leftskip   = 0pt plus1fill
  \rightskip  = 0pt plus1fill
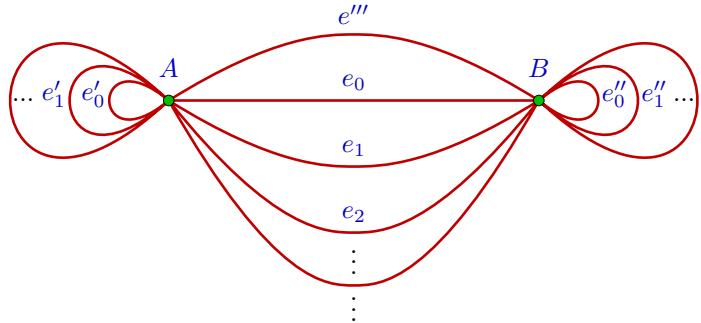
  \noindent Pic.\,3. Equivalent graph for torus.\par}
\par
\vskip1em
\begin{theorem}
  For a decorated graph obtained by attaching a segment to a flat
  $2$-dimensional torus, the following asymptotic estimate holds, as $T$
  goes to infinity:
  $$
    \mfln \mfN(T) \le 3 \left(\frac{5\pi}{8ab} \zeta (3)\right)^{\frac13}T^\frac23 (1+o(1))
  $$
  If $\{t'_i\}_{i=0}^\infty \cup \{t_i\}_{i=0}^\infty$ are linearly
  independent over $\mathbb{Q}$, then inequality turns into equality.
  \label{th3}
\end{theorem}
\indent
  \textbf{Proof.}
  1) Packets reach point $A$ at times of the form
  $$
    W =\Big\{ t''' n''' +
                \sum_{i=0}n'_i t'_i +
                \sum_{i=0} n_i t_i \
              \Big|
              \ n''' + \sum_{i=0}n_i \hbox{ is even} \Big\}.
  $$
  But it may happen that
  at such time another packet comes to the point $A$ by edge $e'''$ All
  such moments are defined by condition $n''' > 0$, and $\sum_{i=0}n_i $
  is even. We exclude from $W$ all such times and obtain
  $$
    W' =\Big\{ \sum_{i=0}n'_i t'_i + \sum_{i=0} n_i t_i \ \Big|
                \ \sum_{i=0}n_i \hbox{ is even} \Big\}.
  $$
  At each time from $W'$ a new packet starts from point $A$ by edge $e'''$.
\par
\indent
  2) Similarly we obtain that at each time from
  $$
    W'' =\Big\{ \sum_{i=0}n'_i t'_i + \sum_{i=0} n_i t_i \ \Big|
                \ \sum_{i=0}n_i \hbox{ is odd} \Big\}
  $$
  a new packet starts from point $B$ by edge $e'''$.
\par
\indent
  We join times from cases 1 and 2: $Q = W' \cup W''$
  $$
    Q = \Big\{\sum_{i=0}n'_i t'_i + \sum_{i=0} n_i t_i \ \Big\}
  $$
\par
\indent
  Thus $\mfN(T) = \{ t\in Q| t \le T \}$.
\par
\indent
  If $\{t'_i\}_{i=0}^\infty \cup \{t_i\}_{i=0}^\infty$ are linearly
  independent over $\mathbb{Q}$, then there is a bijection between $Q$ and
  sets $\{n_i\}_{i=0}^\infty \cup \{n'_i\}_{i=0}^\infty$. Thus $\mfN(T)$ is
  equal to the number of inequality solutions $\sum_{i=0}n'_i t'_i +
  \sum_{i=0} n_i t_i \le T$.
\par
\indent
  Counting function in our situation is given by the following formula
  $$
    \rho(\lambda) = \# \{ \sqrt{(c+na)^2+(d+mb)^2} \le \lambda| n,m\in
      \mathbb{Z} \} +
  $$
  $$
    \# \{ \sqrt{(na)^2+(mb)^2} \le \lambda | n,m\ge 0,
    n,m\in \mathbb{Z} \} = \frac{5\pi \lambda^2}{4ab} (1+
    O(\lambda^{-\varepsilon}))
  $$
\par
\indent
  We apply \ref{th1} with $c_0 = \frac{5\pi}{4ab}, \gamma=1$ and finish
  the proof.
\par
\indent
  If $\{t'_i\}_{i=0}^\infty \cup \{t_i\}_{i=0}^\infty$ are linearly
  dependent over $\mathbb{Q}$ then different sets of integers
  $\{n_i\}_{i=0}^\infty \cup \{n'_i\}_{i=0}^\infty$ can correspond to one
  time of packets birth. It means that Theorem \ref{th1} gives an upper
  bound for $\mfln \mfN(T)$.

\par
\indent
  Let us consider a decorated graph obtained by gluing an edge $e'''$ to a
  3-dimensional flat torus with fundamental cycle lengths equal $a, b, c$.
  Suppose that edge is glued at points $A=(0,0,0), B=(d,e,f)$. In this
  case
  $$
    \{t_i\}_{i=0}^\infty = \{
    \sqrt{(d+na)^2+(e+bm)^2+(f+cl)^2}|n,m,l \in \mathbb{Z} \}
  $$
  $$
    \{t'_i\}_{i=0}^\infty = \{ \sqrt{(na)^2+(bm)^2+(cl)^2}|n,m,l \ge 0,
    n,m,l \in \mathbb{Z} \}.
  $$
\par
\indent
  Similarly to the previous theorem we can obtain
\par
\begin{theorem}
  For a decorated graph obtained by attaching a segment to
  a flat 3-dimensional torus, the following asymptotic estimate holds, as
  $T$ goes to infinity:
  $$
    \mfln \mfN(T) \le 4 \left(\frac{\pi}{3abc} \zeta
    (4)\right)^{\frac14} T^\frac34 (1+o(1))
  $$
  If $\{t'_i\}_{i=0}^\infty \cup \{t_i\}_{i=0}^\infty$ are linearly
  independent over $\mathbb{Q}$ then inequality turns into equality.
  \label{th4}
\end{theorem}
  \relax

   \section{Uniformly secure manifolds}
  \label{USM}
\noindent

{\bfseries Definition.} A Riemannian manifold is said to be \emph{uniformly secure}
if there is a finite number $s$ such that all geodesics connecting an
arbitrary pair of points in the manifold can be blocked by $s$ point
obstacles.

Let us remind a theorem that describes uniformly secure manifolds.

\begin{theorem} (K. Burns, E. Gutkin). Let $M$ be a compact Riemannian manifold that is uniformly
secure. Then the topological entropy of the geodesic flow for $M$ is zero, and the fundamental group of $M$ is virtually nilpotent.
If, in addition, $M$ has no conjugate points, then $M$ is flat.
\end{theorem}

We prove a theorem for decorated graphs constructed from uniformly secure manifolds.

\bigskip

\begin{theorem}
Let a segment with the travel time $L$ be glued at two points $A$ and $B$ on the surface $M$.
Suppose that for geodesics connecting $A$ with $A$, $B$ with $B$, $A$ with $B$ the following condition holds: the number $g(\lambda)$ of geodesics whose length is equal or less than $\lambda$ equals
$$
g(\lambda) = c_0\lambda^{1+ \gamma}(1 + O(\lambda^{-\varepsilon})), \varepsilon> 0.
$$
Then
$$
ln N(T) \le (\gamma + 2) \left(\frac {3 c_0 \Gamma(\gamma + 2)\zeta(\gamma + 2)}{(\gamma + 1)^{\gamma + 1}}\right)^{\frac {1}{\gamma + 2}}T^{\frac {\gamma + 1}{\gamma + 2}}(1 + o (1)).
$$
\end{theorem}

\textbf{Proof} We denote by $t'_i (i \ge 0)$ the length (i.e. propagation time) of geodesics joining $A$ and $A$, by $t_i (i \ge 0)$ the length of the geodesic connecting $A$ and $B$, by $t''_i (i \ge 0)$ the length of the geodesic connecting $B$ and $B$.

Let the packet came from point A. $N (T)$ can be increased only in the time of the form (each of these times corresponds to a path from $A$ to $A$ or $A$ to $B$)
$$
T = \left \{\sum t'_i n'_i + \sum t'' _ i n'' _ i + \sum t_i n_i + Ln''' \right \}
$$
where $n''', n_i, n'_i, n''_ i$ are non-negative integers such that if $\forall i: n_i = 0$ and $ n'''= 0$, then $\forall i: n''_ i = 0$.
Let $T_A$ be the times in which packets arrive at the point $A$. Elements of $T_A$ are characterized by the condition: $n'' + \sum n_i$ is even.
Similarly, let $T_B$ be times in which packets arrive at the point $B$. Elements of $T_B$ characterized by the condition: $n '' + \sum n_i$ is odd.

Then $T = T_A \cup T_B$.
Now we decompose $T_A = T_ {A_1} \cup T_ {A_2}$, where $T_ {A_1}$ are times satisfying $n'''\ne 0$, and $T_ {A_2}$ are times satisfying condition $n '''= 0$. Similarly, we decompose $T_B = T_ {B_1} \cup T_ {B_2}$, where $T_ {B_1}$ are times satisfying $n'''\ne 0$, and $T_ {B_2}$ are times satisfying condition $n'''= 0$.

Note that the number of packets on the edge $e'''$ will not increase in the time belongs to $T_ {A_1}$ or $T_ {B_1}$ because always there exists packets that come from the edge $e'''$ at this time.

Therefore, $N (T)$ does not exceed the number of solutions of the inequality $\sum t'_i n'_i + \sum t'' _ i n'' _ i + \sum t_i n_i \le T$. Analogically, $N(T)$ does not exceed the number of solutions of the inequality $\sum s_i n_i \le T$.

\begin{con}
Let a segment with the travel time $L$ be glued at two points $A$ and $B$ on the surface $M$.
Suppose that $M$ is uniformly secure.
Then
  $$
    \ln N(T) \le
    (q + 2) \left( \frac{3C_g \Gamma (q + 2) \zeta(q +
    2)}{(q + 1)^{q + 1}} \right)^{\frac{1}{q + 2}}
    T^{\frac{q + 1}{q + 2}} (1+o(1)),
 $$
where $C_g$ and $q$ depends only on $M$.
\label{usm_th}
\end{con}

\textbf{Proof}
We use the following theorem (see \cite{BuGu}):

\begin{lemma} (K. Burns, E. Gutkin). Let $M$ be a compact Riemannian manifold. If $M$ is uniformly secure, then there are positive constants $C$ and $q$ such that for any pair $x, y \in M$ we have
$CF_{(x, y)}(T){\leq}C_gT^q$.
\end{lemma}

So the number of geodesics grows polynomially. Then we apply our previous theorem and get the result.

  \section{Further questions}
\noindent
  We have obtained estimates for $\mfln \mfN(T)$ for the situation where the number of geodesics grows polynomially. Such  surfaces are not only the cylinder and flat tori. For all manifolds with
  polynomial growth it is possible to carry out similar calculations and obtain an upper bound for the number of Gaussian
  packets.
\par
\indent
  But, as already mentioned above, for a large class of Riemannian
  manifolds the number $CF_T(x,y)$ grows as $e^{hT}$. In this case
  Theorem \ref{th1} is not applicable and the question of the asymptotic
  behavior of $\mfN(T)$ remains open. Moreover, the arithmetic properties
  of geodesic lengths affect the accuracy of the estimate. Therefore, it
  would be interesting to consider results related to the linear
  independence over $\mathbb{Q}$ of the lengths of geodesics joining two
  given points on a surface.
\par
  \relax
  \section{Acknowledgments}
\noindent
The authors are grateful to A.\,I. Shafarevich, N.\,S. Gusev and O.\,V. Sobolev, V.\,E. Nazaikinskii and N.\,G. Moschevitin,
for useful discussions and attention to their work. The work was
supported by the grant ``The National Research University Higher School
of Economics' Academic Fund Program in 2013-2014, research grant
No.12-01-0164''. Authors thank Joseph H. Silverman and Jan-Christoph Schlage-Puchta for their very useful comments.

\par
  \relax
  
\end{document}